\documentclass[12pt,eqno,epsf]{article}
\usepackage{amsmath,amssymb,graphicx,slashbox}
\numberwithin{equation}{section}

\def\mydate{January 22, 2011}
\def\ignore#1{{}}

\newcounter{sxn}

\newcounter{axn}

\date{}

\newdimen\mybaselineskip
\renewcommand{\thefootnote}{\arabic{footnote}}

\newcommand{\beeq}{\begin{equation}}
\newcommand{\eneq}{\end{equation}}
\newcommand{\beqn}{\begin{eqnarray}}
\newcommand{\eeqn}{\end{eqnarray}}

\newcommand{\alp}{\alpha}
\newcommand{\bt}{\beta}
\newcommand{\gm}{\gamma}
\newcommand{\Gm}{\Gamma}
\newcommand{\dlt}{\delta}

\newcommand{\vep}{\varepsilon}
\newcommand{\tht}{\theta}

\newcommand{\kp}{\kappa}
\newcommand{\lmd}{\lambda}
\newcommand{\Lmd}{\Lambda}
\newcommand{\sgm}{\sigma}

\newcommand{\vph}{\varphi}
\newcommand{\omg}{\omega}

\newcommand{\be}{\begin{equation}}
\newcommand{\ee}{\end{equation}}
\newcommand{\bea}{\begin{eqnarray}}
\newcommand{\eea}{\end{eqnarray}}
\newcommand{\eql}{\!\!\!&=\!\!\!&}

\newcommand{\sma}{\!\!\!&\simeq\!\!\!&}
\newcommand{\defa}{\!\!\!&\equiv\!\!\!&}

\newcommand{\simgt}{\stackrel{>}{{}_\sim}}
\newcommand{\simlt}{\stackrel{<}{{}_\sim}}

\newcommand{\tl}[1]{\tilde{#1}}
\newcommand{\bdm}[1]{{\mbox{\boldmath $#1$}}}
\newcommand{\tr}{{\rm tr}}

\newcommand{\diag}{{\rm diag}}
\newcommand{\der}{\partial}
\newcommand{\dr}{\!\!d}
\newcommand{\hc}{{\rm h.c.}}
\newcommand{\ie}{{\it i.e.}}

\newcommand{\vev}[1]{\langle #1 \rangle}

\newcommand{\brkt}[1]{\left( #1 \right)}
\newcommand{\brc}[1]{\left\{ #1 \right\}}
\newcommand{\sbk}[1]{\left[ #1 \right]}
\newcommand{\abs}[1]{\left| #1 \right|}

\renewcommand{\Re}{{\rm Re}\,}

\newcommand{\cD}{{\cal D}}

\newcommand{\cI}{{\cal I}}
\newcommand{\cK}{{\cal K}}
\newcommand{\cL}{{\cal L}}
\newcommand{\cM}{{\cal M}}

\newcommand{\cO}{{\cal O}}

\newcommand{\cR}{{\cal R}}

\newcommand{\cph}{c_\phi}
\newcommand{\sph}{s_\phi}

\newcommand{\suL}{SU(2)_{\rm L}}
\newcommand{\suR}{SU(2)_{\rm R}}
\newcommand{\uy}{U(1)_Y}
\newcommand{\uem}{U(1)_{\rm EM}}
\newcommand{\mKK}{m_{\rm KK}}
\newcommand{\thH}{\theta_{\rm H}}
\newcommand{\kL}{\varphi}
\newcommand{\rd}{r}

\begin{document}
\thispagestyle{empty}

\baselineskip=12pt

{\small \noindent \mydate    
\hfill }

{\small \noindent \hfill  KEK-TH-1402}

\baselineskip=35pt plus 1pt minus 1pt

\vskip 1.5cm

\begin{center}
{\Large \bf Radion and Higgs masses }
{\Large \bf in gauge-Higgs unification}\\

\vspace{1.5cm}
\baselineskip=20pt plus 1pt minus 1pt

\normalsize

{\bf Yutaka\ Sakamura}$\!${\def\thefootnote{\fnsymbol{footnote}}
\footnote[1]{\tt e-mail address: sakamura@post.kek.jp}}

\vspace{.3cm}
{\small \it KEK Theory Center, Institute of Particle and Nuclear Studies, 
KEK, \\ Tsukuba, Ibaraki 305-0801, Japan} \\ \vspace{3mm}
{\small \it Department of Particles and Nuclear Physics, \\
The Graduate University for Advanced Studies (Sokendai), \\
Tsukuba, Ibaraki 305-0801, Japan} 
\end{center}

\vskip 1.0cm
\baselineskip=20pt plus 1pt minus 1pt

\begin{abstract}
We evaluate the radion and Higgs masses 
in the gauge-Higgs unification models on the warped geometry, 
in which the modulus is stabilized by the Casimir energy. 
We analyze the one-loop effective potential and 
clarify the dependences of those masses on 
the Wilson line phase~$\thH$. 
The radion mass varies 1-30~GeV for $0.06\leq\sin\thH\leq 0.3$, 
while the Higgs mass is 150-200~GeV 
and depends on $\thH$ only logarithmically. 
The radion couplings to the standard model particles are sensitive to  
the warp factor, and are too small to detect at colliders 
in the region where the five-dimensional description is valid. 
\end{abstract}


\newpage

\section{Introduction}
The gauge-Higgs unification scenario is an interesting candidate for 
the physics beyond the standard model, which was originally proposed 
in Refs.~\cite{Fairlie:1979at,Hosotani:1983xw} 
and revived by Refs.~\cite{Hatanaka:1998yp,Pomarol:1998sd} 
as a solution to the naturalness problem. 
In this class of models, the Higgs mass is protected 
against large radiative corrections 
thanks to a higher-dimensional gauge symmetry~\cite{Antoniadis:2001cv}. 
The models are characterized by the Wilson line phase 
along the extra dimension~$\thH$. 
The electroweak symmetry breaking occurs 
when $\sin\thH\neq 0$ or $\sin\frac{\thH}{2}\neq 0$, 
depending on the models. 
The fluctuation around the vacuum expectation value (VEV) of $\thH$ 
corresponds to the physical Higgs boson in the standard model. 

This scenario has been first investigated 
in the flat spacetime~\cite{Csaki:2002ur,Hall:2001zb}, 
and extended to the Randall-Sundrum warped spacetime~\cite{Randall:1999ee}. 
The models in the latter can solve some problems that exist in the former case. 
The masses of the Higgs and the Kaluza-Klein (KK) modes are enhanced by 
a logarithm of the large warp factor~\cite{Hosotani:2005nz} 
so that they can evade the experimental lower bounds, 
and the large top quark mass can easily be realized 
only by the localization of the mode functions 
in the extra dimension~\cite{Agashe:2004rs}. 
Furthermore, such models have phenomenologically interesting 
features~\cite{Agashe:2004rs}-\cite{Hosotani:2009jk}. 
Hence, we will focus on the Randall-Sundrum spacetime 
as a background geometry in this paper. 

When we work in extra-dimensional models, the stabilization mechanism 
for the size of the extra dimension, which is often called the modulus 
or the radion, must be considered. 
One of the simplest mechanisms 
for the modulus stabilization is proposed in Ref.~\cite{Goldberger:1999uk}. 
A five-dimensional (5D) bulk scalar field plays an essential role 
for the stabilization in this mechanism. 
The modulus can also be stabilized 
by the Casimir energy of the bulk fields. 
This possibility has been discussed in many 
papers~\cite{Fabinger:2000jd}-\cite{Ponton:2001hq},  
and it has been shown that 5D gauge and fermion fields 
that spread over the bulk are essential 
for the modulus stabilization~\cite{Garriga:2002vf}. 
Thus the latter mechanism is more economical in the gauge-Higgs unification scenario 
because the bulk gauge and fermion fields already exist in the theory 
and no extra bulk scalar fields need not be introduced 
just for the stabilization. 

In our previous work~\cite{Maru:2010ap}, we discussed 
the modulus stabilization by the Casimir energy 
in the model proposed in Ref.~\cite{Hosotani:2008tx}, 
in which the Wilson line phase is dynamically determined 
as $\thH=\frac{\pi}{2}$. 
We found there that the brane kinetic terms for the gauge fields 
are necessary for the modulus stabilization, and the radion mass 
is $\cO(1~\mbox{GeV})$. 
Although this model has phenomenologically interesting 
features~\cite{Hosotani:2009jk}, 
the electroweak precision measurements disfavor $\thH=\frac{\pi}{2}$ 
according to the analysis in Ref.~\cite{Agashe:2004rs}. 
Besides, it is a nontrivial task to clarify 
the $\thH$-dependence of the radion and Higgs masses 
because 
the effective potential~$V_{\rm eff}$ 
depends on parameters that control the VEV of $\thH$ in a complicated way. 
Therefore, in this paper, we will extend our previous work~\cite{Maru:2010ap} 
to the case that $\thH$ can take small values and 
clarify the $\thH$-dependence of the radion and Higgs masses 
by evaluating $V_{\rm eff}$. 
We will also discuss the experimental constraints on the radion mass. 

The paper is organized as follows. 
In the next section, we provide a brief review of the model 
in Ref.~\cite{Hosotani:2008tx} focusing on the matter sector, 
and show the one-loop effective potential for the radion 
and Higgs fields. 
In Sec.~\ref{model_extensions}, we extend the matter sector of the model 
to realize small values of $\thH$, and see how the effective potential 
is modified by such extensions. 
In Sec.~\ref{Rad_stb}, we estimate 
the radion and Higgs masses as functions of $\thH$, 
and comment on the experimental constraints on the radion mass.  
Sec.~\ref{summary} is devoted to the summary. 
We define some functions useful for our analysis 
in Appendix~\ref{spectrum}, show an approximate form of 
the effective potential in Appendix~\ref{ap:Veff}, and provide some 
useful expressions for the numerical calculation 
in Appendix~\ref{expr_for_cal}.

\section{$\bdm{SO(5)\times U(1)_X}$ model}
In this paper, we consider the gauge-Higgs unification models 
based on a 5D $SO(5)\times U(1)_X$ gauge theory.  
This class of models was first discussed in Ref.~\cite{Agashe:2004rs}, and  
several similar models with different matter sectors have been 
studied so far~\cite{Contino:2006qr,Hosotani:2008tx,Medina:2007hz}. 
In our previous work~\cite{Maru:2010ap}, we considered 
a model proposed in Ref.~\cite{Hosotani:2008tx} as the simplest example. 
We start with a brief review of this model, focusing on the matter sector, 
and extend it later. 

We assume the 5D warped spacetime compactified 
on an orbifold~$S^1/Z_2$~\cite{Randall:1999ee} as a background geometry. 
The background metric is given by 
\be
 ds^2 = G_{MN}dx^Mdx^N = e^{-2\sgm(y)}\eta_{\mu\nu}dx^\mu dx^\nu+dy^2, 
\ee
where $M,N=0,1,2,3,4$ are 5D indices and $\eta_{\mu\nu}=\diag(-1,1,1,1)$. 
The fundamental region of $S^1/Z_2$ is $0\leq y\leq L$. 
The function~$e^{\sgm(y)}$ is a warp factor, and 
$\sgm(y)=ky$ in the fundamental region, where $k$ is 
the inverse AdS curvature radius. 
The orbifold has two fixed points~$y=0$ and $y=L$, which are called 
the UV and IR branes, respectively. 
The gauge symmetry is broken to $\suL\times\suR\times U(1)_X$ at the IR brane, 
and to $\suL\times\uy$ at the UV brane 
by boundary conditions~\cite{Agashe:2004rs}. 
In order to stabilize the modulus, we need 
brane-localized kinetic terms for the gauge fields~\cite{Garriga:2002vf}. 
Thus we introduce the following terms on the IR brane. 
\be
 \cL_{\rm bd}^{\rm kin} = 2\sqrt{-g}\sbk{
 -\frac{\kp_c}{4k}\tr\brc{F_{\mu\nu}^{(G)}F^{(G)\mu\nu}}
 -\frac{\kp_w}{4k}\tr\brc{F_{\mu\nu}^{(A)}F^{(A)\mu\nu}}
 -\frac{\kp_x}{4k}F_{\mu\nu}^{(B)}F^{(B)\mu\nu}}\dlt(y-L), 
 \label{L_bd^kin}
\ee
where $\sqrt{-g}\equiv\det(g_{\mu\nu})$, $g_{\mu\nu}$ is 
the 4D induced metric on the IR brane, 
$F_{MN}^{(G)}$, $F_{MN}^{(A)}$, $F_{MN}^{(B)}$ are field strengths 
for the $SU(3)_C$, $SO(5)$, $U(1)_X$ gauge fields, 
and $\kp_c$, $\kp_w$, $\kp_x$ are dimensionless constants. 
For simplicity, we do not consider kinetic terms on the UV brane 
nor brane kinetic terms for the 5D fermions, and assume that 
$\kp\equiv\kp_c=\kp_w=\kp_x$ in the following.

\subsection{Matter sector}
We introduce 5D fermions~$\Psi_i$ ($i=1,2,\cdots$) belonging to 
the vectorial representation of $SO(5)$ as matter fields. 
The 5D Lagrangian in this sector is given by 
\be
 \cL = \sqrt{-G}\sbk{\sum_i\brc{i\bar{\Psi}_i\Gm^N\cD_N\Psi_i
 -iM_{\Psi i}\vep(y)\bar{\Psi}_i\Psi_i}+\cdots}, 
\ee
where $G\equiv\det(G_{MN})$, $\Gm^N$ are 5D gamma matrices 
contracted by the f\"{u}nfbein, $\cD_N$ is the covariant derivative, 
and $\vep(y)$ is a periodic step function. 
The ellipsis denotes the gauge sector. 

It is useful to express the $SO(5)$ vector $\Psi=(\psi_1,\cdots,\psi_5)^t$ as 
\be
 \Psi = \sbk{\begin{pmatrix} \hat{\psi}_{11} \\ \hat{\psi}_{21} \end{pmatrix}, 
 \; \begin{pmatrix} \hat{\psi}_{12} \\ \hat{\psi}_{22} \end{pmatrix}, \;
 \psi_5}, 
\ee
where 
\be
 \hat{\psi} = \begin{pmatrix} \hat{\psi}_{11} & \hat{\psi}_{12} \\
 \hat{\psi}_{21} & \hat{\psi}_{22} \end{pmatrix} 
 \equiv \frac{1}{\sqrt{2}}\brkt{\psi_4\bdm{1_2}+i\vec{\psi}\cdot\vec{\sgm}}
 i\sgm_2 
\ee
is a bidoublet and $\psi_5$ is a singlet for $\suL\times\suR$. 
For example, the third generation of the quark sector comes from 
two multiplets~$\Psi_1$ and $\Psi_2$, which are expressed as 
\bea
 \Psi_1 \eql \sbk{Q_1=\begin{pmatrix} T \\ B \end{pmatrix}, \;
 q=\begin{pmatrix} t \\ b \end{pmatrix}, \; t'}, \nonumber\\
 \Psi_2 \eql \sbk{Q_2=\begin{pmatrix} U \\ D \end{pmatrix}, \;
 Q_3=\begin{pmatrix} X \\ Y \end{pmatrix}, \; b'}. 
\eea
The orbifold parities for them are listed in Table~\ref{Z2_parity}. 
The subscript~$R$ denotes the 4D right-handed chirality 
defined by $\gm_5\equiv\Gm^4$. 
The left-handed components have the opposite parities 
to the right-handed ones. 
\begin{table}[t,b]
\begin{center}
\begin{tabular}{|c|c|c||c|c|c|}
 \hline \rule[-2mm]{0mm}{7mm} 
 $Q_{1R}$ & $q_{R}$ & $t'_{R}$ & 
 $Q_{2R}$ & $Q_{3R}$ & $b'_{R}$ \\ \hline 
 $(-,-)$ & $(-,-)$ & $(+,+)$ &  
 $(-,-)$ & $(-,-)$ & $(+,+)$  \\ \hline
\end{tabular}
\end{center}
\caption{The orbifold parities for the fermions. 
The left and the right signs in the parenthesis denote 
parities around $y=0$ and $y=L$, respectively. 
The left-handed components have the opposite parities 
to the right-handed ones. 
}
\label{Z2_parity}
\end{table}
On the UV brane, we can introduce brane-localized chiral fermion 
fields and change the boundary conditions there, 
just like we did in Ref.~\cite{Hosotani:2008tx}. 
The resulting boundary conditions on the UV brane are 
\bea
 &&Q_{1L} = c_\omg q_{L}+s_\omg Q_{2L} = Q_{3L} = 0, \nonumber\\
 &&-s_\omg q_{R}+c_\omg Q_{2R} = 0, \;\;\;\;\;
 t'_{L} = b'_{L} = 0,  
%
 \label{UVbdcd1} 
\eea
at $y=0$. 
Here $s_\omg\equiv\sin\omg$, $c_\omg\equiv\cos\omg$, 
and $\omg$ is a mixing angle determined by the ratio of 
the boundary mass parameters. 
(Since $q_L$ and $Q_{2L}$ have the same quantum numbers 
for $\suL\times\uy$, they can mix on the UV brane.)

\subsection{Mass spectrum}
The mass spectrum~$\brc{m_n}$ in the 4D effective theory 
is determined as solutions to the equation, 
\be
 \rho^I(\lmd_n;\thH) = 0, \label{mass_det}
\ee
where $I$ specifies the sectors, and $\lmd_n\equiv m_n/k$. 
The functions~$\rho^I(\lmd;\thH)$ are written in terms of the Bessel functions, 
and listed in Appendix~A of Ref.~\cite{Maru:2010ap}.\footnote{
They are denoted as $\rho_I(\lmd)$ in the notation of Ref.~\cite{Maru:2010ap}, 
and we need to generalize the expressions there  
to the case that $M_{\Psi 1}\neq M_{\Psi 2}$ for the quark sector. }
(See (\ref{eg:rho}), for example.)

The W and Z boson masses are obtained as the lightest solutions 
to $\rho^W(\lmd_W;\thH)=0$ and $\rho^Z(\lmd_Z;\thH)=0$, 
and are approximately expressed as 
\bea
 m_W \eql k\lmd_W \simeq \frac{ke^{-kL}\sin\thH}{\sqrt{kL+\kp}}, 
 \nonumber\\
 m_Z \eql k\lmd_Z \simeq \sqrt{\frac{1+\sph^2}{kL+\kp}}ke^{-kL}\sin\thH. 
 \label{ap:mWZ}
\eea
Here, $\sph \equiv g_B/\sqrt{g_A^2+g_B^2}$,  
where $g_A$ and $g_B$ are the 5D gauge couplings 
for the $SO(5)$ and $U(1)_X$ gauge fields. 

\ignore{
The masses of the top and bottom quarks are obtained 
as the lightest solutions 
to $\rho^{2/3}(\lmd_t;\thH)=0$ and $\rho^{-1/3}(\lmd_b;\thH)=0$, respectively. 
When $M_{\Psi 1}\leq M_{\Psi 2}<k/2$, 
their approximate expressions are written as 
\bea
 m_t \eql k\lmd_t \simeq \frac{k\sqrt{1-4c_1^2}\sin\thH}{\sqrt{2}e^{kL}} 
 \simeq \sqrt{\frac{(1-4c_1^2)(kL+\kp)}{2}}m_W, 
 \nonumber\\
 m_b \eql k\lmd_b \simeq \sqrt{\frac{2c_2+1}{2c_1+1}}\frac{c_\omg}{s_\omg}
 e^{(c_1-c_2)kL}m_t,  
 \label{ap:mtb}
\eea
where $c_1\equiv M_{\Psi 1}/k$ and $c_2\equiv M_{\Psi 2}/k$. 
}

\subsection{Effective potential} \label{V_eff}
The one-loop effective potential~$V_{\rm eff}$ for the radion and Higgs fields 
is calculated by the technique in Ref.~\cite{Garriga:2000jb}. 
As mentioned in Ref.~\cite{Garriga:2002vf}, 
only the fields that spread over the bulk 
can give sizable contributions to $V_{\rm eff}$. 
In our model, such fields are 
the gauge fields and the quark multiplets in the third generation.\footnote{ 
A fermion field spreads over the bulk when its bulk mass 
is close to $k/2$. }

Now we promote dimensionless constants~$kL$ and $\thH$ 
to 4D dynamical fields~$\kL(x)$ and $\thH(x)$. 
Then $V_{\rm eff}$ is expressed as the following form. 
\be
 V_{\rm eff}(\kL,\thH) = \frac{k^4}{16\pi^2}\sbk{
 \tau_{\rm UV}+e^{-4\kL}\tau_{\rm IR}+e^{-4\kL}\hat{V}(\kL,\thH)}, 
\ee
where 
\be
 \hat{V}(\kL,\thH) \equiv \sum_I(-)^{2\eta_I}N_I
 \int_0^\infty\dr w\;w^3\ln\frac{\rho^I(iwe^{-\kL};\thH)}
 {\cK^I(we^{-\kL})\cI^I(w)}.  \label{def:hatV}
\ee
Here $\eta_I=0$ ($\frac{1}{2}$) for bosons (fermions), 
$N_I$ is a number of degrees of freedom for a particle in sector~$I$. 
The functions~$\cK^I(w)$ and $\cI^I(w)$ are expressed 
by products of the modified Bessel functions~$e^{-i\alp\pi}K_\alp(w)$ 
and $e^{i\bt\pi}I_\bt^\kp(w)$ respectively,\footnote{
The definition of $K_\bt^\kp(w)$ and $I_\bt^\kp(w)$ are given 
in (\ref{def:IK^kp}). } 
and defined so that $\rho^I(iwe^{-\kL};0)/\cK^I(we^{-\kL})\cI^I(w)$ 
becomes a product of $\brc{1-e^{i(\alp-\bt)\pi}
\frac{I_\alp(we^{-\kL})K_\bt^\kp(w)}{K_\alp(we^{-\kL})I_\bt^\kp(w)}}$. 
(See Appendix~B of Ref.~\cite{Maru:2010ap}.)
The dimensionless constants $\tau_{\rm UV}$ and $\tau_{\rm IR}$ 
cannot be determined in the context of the 5D field theory.\footnote{
The divergent one-loop contributions to them are given 
in (B.10) of Ref.~\cite{Maru:2010ap}. 
They are absorbed in the renormalization of the tensions 
of the UV and IR branes, respectively. }

In our model, $V_{\rm eff}$ is approximately expressed as 
\be
 V_{\rm eff}(\kL,\thH) \simeq V_0(\kL)+V_2(\kL)\cos^2\thH, 
 \label{ap:Veff1}
\ee
where $V_0(\kL)$ and $V_2(\kL)$ are independent of $\thH$. 
From the stationary condition for $\thH$, we obtain 
\be
 \sin 2\thH = 0.  \label{stnry_pt1}
\ee
Thus we have $\thH=0,\pm\frac{\pi}{2},\pi$ as candidates for the vacuum. 
As mentioned in Ref.~\cite{Hosotani:2008tx}, 
the contribution from the gauge sector 
prefer the symmetric phase~$\thH=0,\pi$, while that from 
the fermion sector does the broken phase~$\thH=\pm\frac{\pi}{2}$. 
In fact, due to a large contribution from the top quark sector, 
$\thH=\pm\frac{\pi}{2}$ is selected as a vacuum,  
and the electroweak symmetry is broken.

\section{Extensions of the model} \label{model_extensions}
According to the analysis of Ref.~\cite{Agashe:2004rs}, 
the VEV of $\thH$ must be small, \ie, 
\be
 \sin\thH \simlt 0.3\,\mbox{-}\,0.5, 
\ee
from the constraint on the oblique parameter~$S\simlt 0.3$. 

To realize a small value of $\thH$, the matter sector has to be extended. 
The simplest extension is to introduce an additional fermion 
multiplet~$\Psi_3$ that belongs to the spinorial representation of $SO(5)$ 
and whose $U(1)_X$ charge is $1/6$. 
It is decomposed as 
\be
 \Psi_3 = \sbk{\hat{Q},\; \hat{t},\; \hat{b}}, 
\ee
where $\hat{Q}$, $\hat{t}$ and $\hat{b}$ transform as $\bdm{2_{1/6}}$, 
$\bdm{1_{2/3}}$ and $\bdm{1_{-1/3}}$ under $\suL\times\uy$. 
The orbifold parity of each component is assumed as shown 
in Table~\ref{Z2_parity2}. 
\begin{table}[t,b]
\begin{center}
\begin{tabular}{|c|c|c|}
 \hline \rule[-2mm]{0mm}{7mm} 
 $\hat{Q}_{R}$ & $\hat{t}_{R}$ & $\hat{b}_{R}$ 
 \\ \hline 
 $(+,-)$ & $(-,+)$ & $(-,+)$ 
 \\ \hline
\end{tabular}
\end{center}
\caption{The orbifold parities for the components of $\Psi_3$. 
The left-handed components have the opposite parities to 
the right-handed ones. }
\label{Z2_parity2}
\end{table}
This sector consists of the $Q_{\rm EM}=\frac{2}{3}$ and 
$Q_{\rm EM}=-\frac{1}{3}$ sectors, and 
their mass spectra are determined by (\ref{mass_det}) with 
\bea
 \rho^{\Psi_3(2/3)}(\lmd;\thH) \eql \rho^{\Psi_3(-1/3)}(\lmd;\thH)
 = F_{c_3+\frac{1}{2},c_3+\frac{1}{2}}^0(\lmd)
 F_{c_3-\frac{1}{2},c_3-\frac{1}{2}}^0(\lmd)
 -\frac{4\cos^2\frac{\thH}{2}}{\pi^2\lmd^2e^{kL}},  
 \label{rho_Psi3}
\eea
where the function~$F^\kp_{\alp,\bt}(\lmd)$ is defined 
by (\ref{def:Fs}), and $c_3\equiv M_{\Psi 3}/k$. 
This can easily be obtained in the usual procedure 
to determine the mass spectra in the warped spacetime 
(see, for instance, Ref.~\cite{Hosotani:2006qp}). 
Note that the period of the spectrum is $2\pi$, which is twice of 
those in the other sectors. 
If the bulk mass~$M_{\Psi 3}$ is close to $k/2$, 
a contribution from $\Psi_3$ to $V_{\rm eff}$ is sizable.\footnote{
For the anomaly cancellation, additional fermion multiplets are required. 
However, they are irrelevant to the current discussion 
unless their bulk masses are close to $k/2$. 
} 
The approximate expression of $V_{\rm eff}$ is modified as 
\be
 V_{\rm eff}(\kL,\thH) \simeq V_0(\kL)+V_1(\kL)\cos\thH+V_2(\kL)\cos^2\thH. 
 \label{ap:Veff2}
\ee
Now we have a linear term for $\cos\thH$.\footnote{
$V_0$ is also modified from those in (\ref{ap:Veff1}) 
by the $\Psi_3$-contribution. 
} 
Then, we find a new stationary point of $V_{\rm eff}$, 
\be
 \cos\thH \simeq -\frac{V_1}{2V_2},  \label{stnry_pt2}
\ee
if $\abs{V_1}<2\abs{V_2}$. 
When $V_2(\kL)>0$, this gives a global minimum of $V_{\rm eff}$ 
for a fixed value of $\kL$. 
The value of (\ref{stnry_pt2}) is controlled by the bulk mass~$M_{\Psi 3}$. 
In fact, a small value of $\thH$ can be realized by choosing it 
such that $M_{\Psi 1}\simeq M_{\Psi 3}$. 

There is another extension of the matter sector. 
In Ref.~\cite{Contino:2006qr}, two additional fermion multiplets~$\tl{\Psi}_1$ 
and $\tl{\Psi}_2$ are introduced, which have the same quantum numbers 
as $\Psi_1$ and $\Psi_2$, respectively. 
The orbifold parity at the UV brane for each component 
of $\tl{\Psi}_1$ and $\tl{\Psi}_2$ 
are the same as those of $\Psi_1$ and $\Psi_2$, 
while the parities at the IR brane for the former are opposite 
to those for the latter. 
Then the following mass terms are allowed on the IR brane. 
\bea
 \cL_{\rm bd}^{\rm mass} \eql 2\sqrt{-g}\left\{
 i\zeta_1\brkt{\bar{Q}_{1L}\tl{Q}_{1R}+\bar{q}_L\tl{q}_R}
 +i\xi_1\bar{t'}_R\tl{t}'_L \right. \nonumber\\
 &&\hspace{10mm}\left. 
 +i\zeta_2\brkt{\bar{Q}_{2L}\tl{Q}_{2R}+\bar{Q}_{3L}\tl{Q}_{3R}}
 +i\xi_2\bar{b'}_R\tl{b}'_L+\hc \right\}\dlt(y-L),  \label{Lmass_bd}
\eea
where $\zeta_{1,2}$ and $\xi_{1,2}$ are dimensionless mass parameters, and 
\be
 \tl{\Psi}_1 = \sbk{\tl{Q}_1,\; \tl{q},\; \tl{t}'}, \;\;\;
 \tl{\Psi}_2 = \sbk{\tl{Q}_2,\; \tl{Q}_3,\; \tl{b}'}. 
\ee
In general, $\tl{q}_L$ and $\tl{Q}_{2L}$ can mix with 
$q_L$ and $Q_{2L}$ on the UV brane since they have the same quantum numbers 
although it is not considered in Ref.~\cite{Contino:2006qr} 
for simplicity. 

The boundary mass terms in (\ref{Lmass_bd}) relate $\Psi_1$ and $\Psi_2$ 
with $\tl{\Psi}_1$ and $\tl{\Psi}_2$ through the equations of motion. 
This induces a quartic term for $\cos\thH$ in $V_{\rm eff}$. 
Namely, the approximate form of $V_{\rm eff}$ now become 
\be
 V_{\rm eff}(\kL,\thH) \simeq V_0(\kL)+V_2(\kL)\cos^2\thH+V_4(\kL)\cos^4\thH. 
 \label{ap:Veff3}
\ee
In this case, a solution of 
\be
 \cos^2\thH \simeq -\frac{V_2}{2V_4}  \label{stnry_pt3}
\ee
can be a candidate for the vacuum value. 
In fact, it becomes a global minimum of $V_{\rm eff}$ 
for a fixed value of $\kL$ when $0<-V_2<2V_4$ is satisfied. 

We can also extend the matter sector by introducing 
$SO(5)$ tensor multiplets~\cite{Contino:2006qr,Medina:2007hz} 
with boundary masses among the bulk fields. 
Also in this case, $V_{\rm eff}$ has the form of (\ref{ap:Veff3}).

\section{Modulus stabilization} \label{Rad_stb}
\subsection{Scalar mass matrix}
From the stationary condition for $\kL$, we obtain 
\be
 \tau_{\rm IR} = \frac{\der_{\kL}\hat{V}}{4}-\hat{V}. 
 \label{value_tauIR}
\ee
Since $\tau_{\rm IR}$ cannot be determined within our setup, 
it should be treated as an input parameter. 
In order for the 5D description to be valid, 
the 5D scalar curvature~$\cR_5=-20k^2$ must satisfy the condition 
$\abs{\cR_5}<M_5^2$, where $M_5$ is the 5D Planck mass~\cite{Davoudiasl:1999tf}. 
Namely, $\zeta\equiv M_5/k\simgt 4.5$. 
For a sufficiently large warp factor, this means that 
\be
 k \simeq \sqrt{\frac{2}{\zeta^3}}M_{\rm Pl}, 
\ee
where $M_{\rm Pl}$ is the 4D Planck mass. 
By using (\ref{ap:mWZ}), we obtain 
\be
 e^{kL}\sqrt{kL+\kp} \simeq \frac{M_{\rm Pl}}{m_W}
 \sqrt{\frac{2}{\zeta^3}}\sin\thH 
 \simlt 4.5\times 10^{15}\sin\thH. 
 \label{def:zt}
\ee
This means that the warp factor has an upper bound depending on $\thH$. 
An extremely small value of $\sin\thH$ does not allow 
$e^{kL}\sim 10^{15}$. 
Besides, it requires a fine tuning among the model parameters, 
such as $M_{\Psi i}$ ($i=1,2,3$).  
So we focus on a parameter region where $0.06\leq\sin\thH\leq 0.3$, 
and take the allowed maximal value~$e^{kL}=5\times 10^{13}$ 
in the following analysis. 
From (\ref{value_tauIR}) with the explicit expression of $\hat{V}$, 
we can see that this value of the warp factor is realized 
by an $\cO(1)$ value of $\tau_{\rm IR}$. 
Here (\ref{value_tauIR}) is evaluated at the value of $\thH$ 
determined by (\ref{stnry_pt2}) or (\ref{stnry_pt3}). 
The other constant~$\tau_{\rm UV}$ is determined by the condition 
that the cosmological constant in the 4D effective theory vanishes. 

The bulk masses for the fermions control the profiles of the zero-modes 
and their mass spectrum in the 4D effective theory. 
In the absense of the boundary mixing, 
the zero-mode mass eigenvalue is larger (smaller) than $m_W$ 
when $c<\frac{1}{2}$ ($c>\frac{1}{2}$), 
where $c$ is a ratio of the bulk mass~$M_\Psi$ to $k$. 
(See Fig.~2 and Table~1 in Ref.~\cite{Hosotani:2006qp}.) 
The situation is more complicated in our case 
because of the boundary mixing parametrized by $s_\omg$ 
in (\ref{UVbdcd1}). 
For example, in Ref.~\cite{Hosotani:2008tx}, 
$M_{\Psi 1}=M_{\Psi 2}=0.43k$ and $s_\omg=1-\frac{1}{2}(m_b/m_t)^2$ 
are chosen to reproduce the top and bottom quark masses. 
In this paper, we will choose $M_{\Psi 1}=0.43k$,  
$M_{\Psi 2}=0.53k$ and $s_\omg=0.86$ as an example. 
In the case of (\ref{ap:Veff2}), $M_{\Psi 3}$ determines the value of $\thH$. 

Once the warp factor is given, we can discuss 
the stability of the vacuum. 
By using the stationary conditions, 
the second derivatives of $V_{\rm eff}$ at the minimum are given by 
\bea
 \left.\der_{\kL}^2V_{\rm eff}\right|_0 \eql \frac{k^4e^{-4kL}}{16\pi^2}
 \left.\brkt{\der_{\kL}^2\hat{V}-4\der_{\kL}\hat{V}}\right|_0, \nonumber\\
 \left.\der_{\kL}\der_{\thH}V_{\rm eff}\right|_0 
 \eql \frac{k^4e^{-4kL}}{16\pi^2}\left.\der_{\kL}\der_{\thH}\hat{V}\right|_0, 
 \nonumber\\
 \left.\der_{\thH}^2V_{\rm eff}\right|_0 
 \eql \frac{k^4e^{-4kL}}{16\pi^2}\left.\der_{\thH}^2\hat{V}\right|_0, 
\eea 
where the symbol~$|_0$ indicates that the quantity is evaluated at the minimum 
of $V_{\rm eff}$. 
These provide the mass matrix for the 
fluctuations~$\tl{\kL}\equiv\kL-kL$ and 
$\tl{\tht}_{\rm H}\equiv \thH-\vev{\thH}$. 
Note that we have to canonically normalize these fluctuations  
in order to discuss the physical masses. 
The canonical normalization for them are given by 
\bea
 \rd(x) \eql \sqrt{\frac{3M_5^3}{k(e^{2kL}-1)}}\tl{\kL}, \nonumber\\
 h(x) \eql \sqrt{\frac{4k}{g_A^2(e^{2kL}-1)}}\tl{\tht}_{\rm H}(x). 
\eea
Then the (squared) mass matrix for $\rd$ and $h$ is calculated as 
\be
 \cM_{\rm scalar}^2 = \begin{pmatrix} m_{\rd\rd}^2 & m_{\rd h}^2 \\
 m_{\rd h}^2 & m_{hh}^2 \end{pmatrix}, 
\ee
where
\bea
 m_{\rd\rd}^2 \defa \frac{k(e^{2kL}-1)}{3M_5^3}
 \left.\der_{\kL}^2V_{\rm eff}\right|_0 
 \simeq \frac{k^5e^{-2kL}}{48\pi^2M_5^3}\left.\brkt{
 \der_{\kL}^2\hat{V}-4\der_{\kL}\hat{V}}\right|_0, 
 \nonumber\\
 m_{\rd h}^2 \defa \frac{g_A(e^{2kL}-1)}{2\sqrt{3M_5^3}}
 \left.\der_{\kL}\der_{\thH}V_{\rm eff}\right|_0 
 \simeq \frac{g_Ak^4e^{-2kL}}{32\pi^2\sqrt{3M_5^3}}
 \left.\der_{\kL}\der_{\thH}\hat{V}\right|_0, \nonumber\\
 m_{hh}^2 \defa \frac{g_A^2(e^{2kL}-1)}{4k}
 \left.\der_{\thH}^2V_{\rm eff}\right|_0 
 \simeq \frac{g_A^2k^3e^{-2kL}}{64\pi^2}
 \left.\der_{\thH}^2\hat{V}\right|_0. 
 \label{scalar_mass}
\eea

\subsection{$\bdm{\thH}$-dependence of various mass scales}
Now we express each parameter in terms of the 4D ones, \ie, 
$M_{\rm Pl}$, $m_W$ and the 4D $\suL$ gauge coupling~$g_4$. 
First we should note that $k$ is a function of $\thH$ 
through (\ref{ap:mWZ}) since we take $m_W$ as an input parameter. 
Thus $M_5$ and $g_A$ also depend on $\thH$ as 
\bea
 M_5^3 \eql \frac{2k}{1-e^{-2kL}}M_{\rm Pl}^2
 \simeq \frac{2e^{kL}\sqrt{kL+\kp}}{\sin\thH}M_{\rm Pl}^2m_W, \nonumber\\
 g_A \eql \frac{g_4\sqrt{kL+\kp}}{\sqrt{k}} 
 \simeq g_4\brkt{\frac{\sqrt{kL+\kp}\sin\thH}{e^{kL}m_W}}^{1/2}. 
 \label{thH-dependence}
\eea
\ignore{
Thus, 
the squared mass matrix~(\ref{scalar_mass}) is rewritten as 
\bea
 m_{rr}^2 \sma \frac{e^{2kL}(kL+\kp)^2m_W^4}{96\pi^2M_{\rm Pl}^2\sin^4\thH}
 \left.\brkt{\der_{\kL}^2\hat{V}-4\der_{\kL}\hat{V}}\right|_0, \nonumber\\
 m_{rh}^2 \sma -\frac{g_4e^{kL}(kL+\kp)^2m_W^3}
 {32\sqrt{6}\pi^2M_{\rm Pl}\sin^3\thH}
 \left.\der_{\kL}\der_{\thH}\hat{V}\right|_0, \nonumber\\
 m_{hh}^2 \sma \frac{g_4^2(kL+\kp)^2m_W^2}{64\pi^2\sin^2\thH}
 \left.\der_{\thH}^2\hat{V}\right|_0.  
 \label{expr:m_scalar^2}
\eea
}
Let us first consider the case of (\ref{ap:Veff2}).  
As shown in Appendix~\ref{ap:Veff}, $\hat{V}(\kL,\thH)$ has 
the following approximate form for $\kL\gg 1$. 
\be
 \hat{V}(\kL,\thH) = \sum_{n=0}^2\brkt{u_n+\frac{v_n}{\kL}}\cos^n\thH, 
 \label{expd:hatV}
\ee
where $u_n$ and $v_n$ are constants, and  
$u_0=\cO(-5)$, $u_1=u_2=\cO(3)$, $v_0=v_2=\cO(10)$ 
and $v_1=0$. 
Then, by using (\ref{thH-dependence}) and (\ref{expd:hatV}), 
the expressions in (\ref{scalar_mass}) become 
\bea
 m_{rr}^2 \sma \frac{e^{2kL}m_W^4}{24\pi^2M_{\rm Pl}^2\sin^4\thH}
 \brkt{v_0+v_2\cos^2\thH}, \nonumber\\
 m_{rh}^2 \sma \frac{g_4e^{kL}m_W^3}
 {16\sqrt{6}\pi^2M_{\rm Pl}\sin^2\thH}v_2\cos\thH, \nonumber\\
 m_{hh}^2 \sma \frac{g_4^2 kL^2m_W^2}{32\pi^2}
 \brkt{u_2+\frac{v_2}{kL}}.  
 \label{expr:m_scalar^2}
\eea
Here we have assumed that $\kp\ll kL\simeq 32$. 
Since $e^{kL}m_W/M_{\rm Pl}=1.7\times 10^{-3} \ll 1$ for our choice 
of the warp factor, the mixing angle between the radion and the Higgs boson 
is negligible. 
Notice that there is no kinetic mixing between them in our model, 
which originates from the curvature-scalar 
mixing term~$\sqrt{-g}\xi \cR_4(g)H^\dagger H
\dlt(y-L)$~\cite{Giudice:2000av,Csaki:2000zn} 
($\xi$ is a dimensionless parameter, $\cR_4(g)$ is the 4D Ricci scalar 
for the induced metric~$g_{\mu\nu}$ and $H$ is the Higgs doublet), 
because such a term is prohibited by the 5D gauge symmetry 
when $H$ is a part of the 5D gauge field. 

Thus the radion and Higgs masses~$m_{\rm rad}$ and $m_{\rm H}$ are roughly estimated as 
\bea
 m_{\rm rad} \!\!\!&\sim\!\!\!& 
 \frac{e^{kL}m_W^2}{2\sqrt{6}\pi M_{\rm Pl}\sin^2\thH}
 \sqrt{v_0+v_2} 
 \sim \frac{\sqrt{\cO(20)}\times 10^{-2}\mbox{ GeV}}{\sin^2\thH}, \nonumber\\
 m_{\rm H} \!\!\!&\sim\!\!\!& \frac{g_4kL m_W}{4\sqrt{2}\pi}
 \sqrt{u_2+\frac{v_2}{kL}} 
 \sim \sqrt{\cO(3)}\times 90\mbox{ GeV}. 
 \label{ap:m_scalar2}
\eea
We have used that $\cos\thH\sim 1$.  

We obtain a similar result also in the case of (\ref{ap:Veff3}). 
Now $\hat{V}(\kL,\thH)$ is approximated as 
\be
 \hat{V}(\kL,\thH) = \sum_{n=0}^2\brkt{u_{2n}+\frac{v_{2n}}{\kL}}\cos^{2n}\thH, 
\ee
where $u_0=\cO(-5)$, $u_2=\cO(3)$, $u_4=\cO(1)$, $v_0=v_2=\cO(10)$, 
and $v_4=0$. 
Then, $m_{\rm rad}$ and $m_{\rm H}$ are estimated as 
\bea
 m_{\rm rad} \!\!\!&\sim\!\!\!& 
 \frac{e^{kL}m_W^2}{2\sqrt{6}\pi M_{\rm Pl}\sin^2\thH}
 \sqrt{v_0+v_2}
 \sim \frac{\sqrt{\cO(20)}\times 10^{-2}\mbox{ GeV}}{\sin^2\thH}, \nonumber\\
 m_{\rm H} \!\!\!&\sim\!\!\!& \frac{g_4kL m_W}{2\sqrt{2}\pi}\sqrt{u_4}\cos\thH 
 \sim \sqrt{\cO(1)}\times \mbox{180 GeV}. 
 \label{ap:m_scalar3}
\eea 
The radion-Higgs mixing is negligible also in this case. 

The typical KK mass scale~$\mKK$ is estimated as 
\be
 \mKK \equiv \frac{k}{e^{kL}-1} 
 \sim \frac{\sqrt{kL}m_W}{\sin\thH} \simeq \frac{450\mbox{ GeV}}{\sin\thH}. 
 \label{def:mKK}
\ee

\begin{figure}[t]
\centering  \leavevmode
\includegraphics[width=73mm]{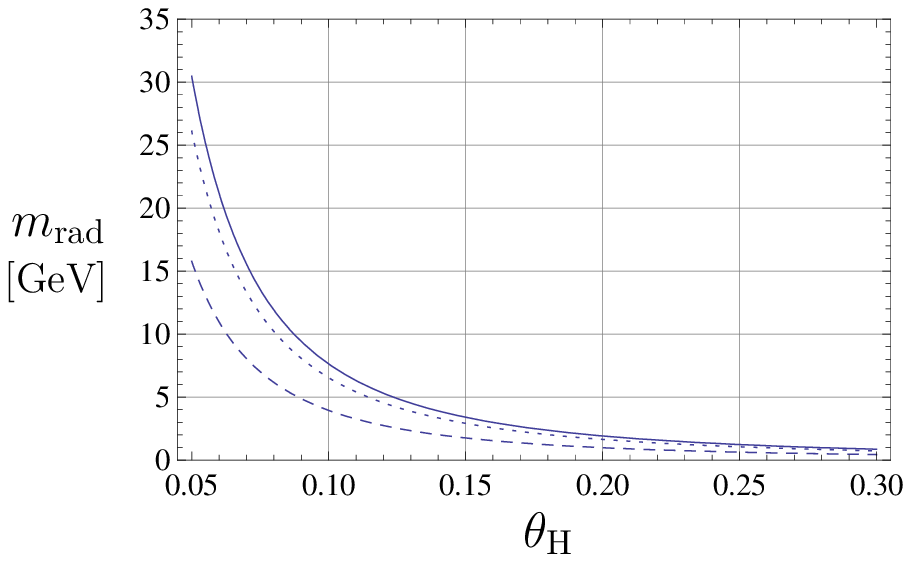} \hspace{10mm}
\includegraphics[width=73mm]{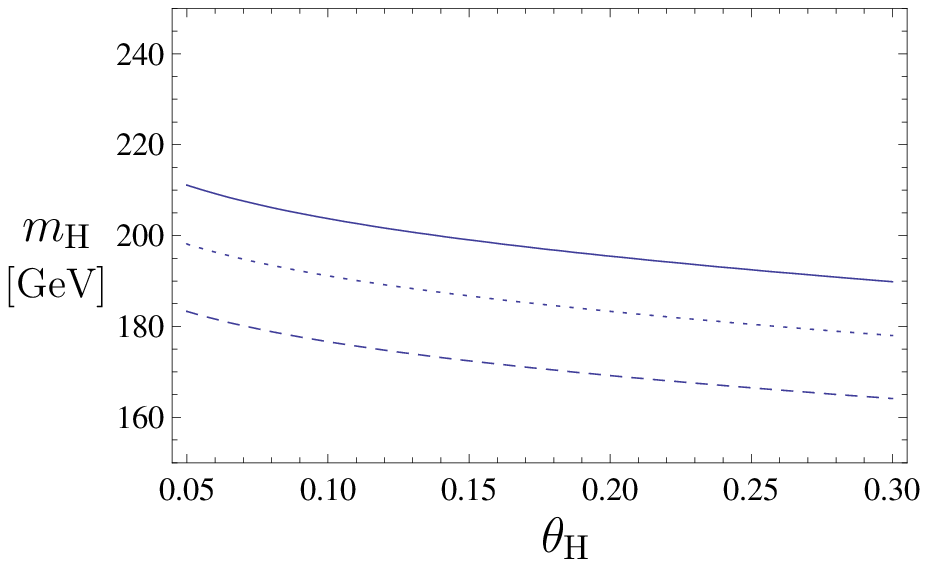}
\caption{The masses of the radion and the Higgs boson as functions of 
$\thH$. 
The solid, dotted, dashed lines represent 
the case of $\kp=5.0,3.0,1.0$, respectively. }
\label{Rad-Hig_mass}
\end{figure}
\begin{figure}[h]
\centering  \leavevmode
\includegraphics[width=70mm]{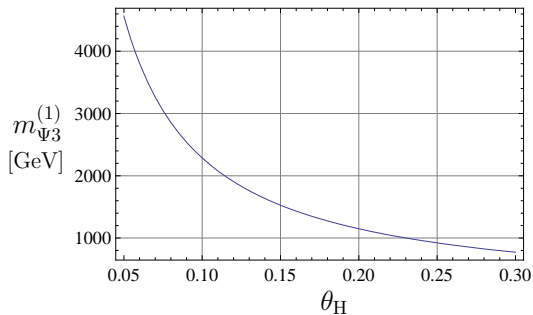} 
\caption{The mass of the first KK mode for $\Psi_3$ 
as a function of $\thH$. }
\label{KK_mass}
\end{figure}
Now we show some numerical results. 
As a specific example, we consider the case of (\ref{ap:Veff2}). 
Fig.~\ref{Rad-Hig_mass} shows the radion and Higgs masses 
as functions of $\thH$.  
The lightest KK mode comes from $\Psi_3$ when $\kp<20$. 
Fig.~\ref{KK_mass} shows its mass as a function of $\thH$. 
From these plots, we can read off the $\thH$-dependence of each mass as 
\bea
 m_{\rm rad} \sma \frac{\mbox{0.04 GeV}}{\sin^2\thH}, \nonumber\\
 m_{\rm H} \sma \brkt{150-10\ln\sin\thH}\mbox{ GeV}, \nonumber\\
 m_{\Psi 3}^{(1)} \sma \frac{k}{2e^{kL}} 
 \simeq \frac{\mbox{230 GeV}}{\sin\thH}, 
\eea
for $\kp=1.0$. 
The radion-Higgs mixing angle is less than $\cO(10^{-4})$ 
for $\sin\thH\geq 0.06$. 
These results are consistent with the rough estimations~(\ref{ap:m_scalar2}) 
and (\ref{def:mKK}). 

Here we comment on the strength of the brane kinetic terms. 
We find that the modulus stabilization requires $\kp\simgt 0.5$. 
For such values of $\kp$, the Higgs mass is larger than 175~GeV 
at $\thH=0.06$. 
Therefore, within the parameter space we consider, there is a region 
that is consistent with 
the latest excluded region~158~GeV$<m_{\rm H}<$175 GeV 
by the Tevatron~\cite{:2010ar}. 
Since the zero-modes for the gauge fields have (at least approximately) 
flat profiles, the 4D and 5D gauge couplings~$g_4$ and $g_5$ are related through 
\be
 \frac{1}{g_4^2}=\frac{L}{g_5^2}\brkt{1+\frac{\kp}{kL}}.  \label{rel:g45}
\ee
Thus the dominant contributions to $g_4$ come from the bulk terms 
and the brane kinetic terms only give small corrections 
for an $\cO(1)$ value of $\kp$. 
On the other hand, the brane kinetic terms with $\cO(1)$ $\kp$ 
affect the masses of the KK modes and their couplings 
to the zero-modes in a sizable way,  
and their impacts are discussed in Ref.~\cite{Maru:2010ap,Davoudiasl:2002ua}, 
for example.

\subsection{Experimental constraints on the radion mass}
Finally we comment on the experimental constraints on the radion mass. 
The couplings of the radion to other fields 
are obtained by expanding the 4D effective action 
in terms of $\tl{\kL}=\rd/\Lmd_{\rd}$, 
where $\Lmd_{\rd}=\sqrt{3M_5^3/(k(e^{2kL}-1))}\simeq\sqrt{6}e^{-kL}M_{\rm Pl}$. 
In the original Randall-Sundrum setup, in which all the standard model 
particles reside on the IR brane, they are expressed 
as~\cite{Csaki:1999mp,Goldberger:1999un} 
\bea
 \cL_{\rm eff} \eql -\frac{1}{\Lmd_{\rd}}\rd T^\mu_{\;\;\mu}+\cdots, 
 \label{Lr_RS}
\eea
where $T_{\mu\nu}$ is the energy-momentum tensor of the standard model.  
Namely, the radion couples to particles on the IR brane just like 
the standard model Higgs does with 
an extra factor~$v/\Lmd_{\rd}$, where $v=246$ GeV. 
To particles propagating in the bulk, the radion couplings deviate 
from (\ref{Lr_RS}), but are of the same order of 
magnitude~\cite{Rizzo:2002pq,Csaki:2007ns}. 
Hence the radion couplings are very weak because $v/\Lmd_{\rd}=2.1\times 10^{-3}$ 
for our choice of the warp factor. 
Thus the radion mass is not constrained from the collider experiments.\footnote{
The radion with a mass 
in the range:~$12~\mbox{GeV}<m_{\rm rad}<90~\mbox{GeV}$ is excluded 
if $(v/\Lmd_{\rd})^2>0.01$~\cite{Barate:2003sz}, 
and it is excluded for $m_{\rm rad}<12~\mbox{GeV}$ if $(v/\Lmd_{\rd})^2>0.1$ 
\cite{Acton:1991pd}. }
It can also be bounded from below by the consideration of 
the neutrino oscillation inside the supernova~\cite{Mahanta:2000mx}, 
but this lower bound is much lower than 1~GeV in our case. 

In addition to the above tree-level couplings, 
the radion couplings to the photons and to the gluons 
also receive sizable one-loop contributions.\footnote{
In contrast to the original Randall-Sundrum model, there are 
tree-level contributions to these couplings in our models 
because the gauge fields propagate in the bulk. }
These couplings are important for the production and the decay processes 
of the radion. 
They can be enhanced compared to the corresponding Higgs 
couplings~\footnote{
The Higgs couplings to the massless gauge bosons are discussed 
in the context of the gauge-Higgs 
unification in Refs.~\cite{Falkowski:2007hz,Maru:2007xn}. 
} times $v/\Lmd_{\rd}$. 
(See, for example, Ref.~\cite{Goldberger:2007zk}.) 
However such enhancements are insufficient for compensating 
the suppression factor~$v/\Lmd_{\rd}$, 
and for discovering the radion at the Large Hadron Collider. 


The situation does not change so much 
even if a larger value of $e^{kL}$ is chosen. 
From the requirement that $M_5/k> 4.5$ and $\sin\thH<0.3$, 
the warp factor is bounded through (\ref{def:zt}) as 
$e^{kL}<2.3\times 10^{14}$ for an $\cO(1)$ value of $\kp$. 
Thus the suppression factor~$v/\Lmd_{\rd}$ cannot be larger than 0.01, 
which is still too small to detect the radion at the colliders. 

In the case that the fermion mass hierarchy is realized by 
the wavefunction localization in the extra dimension, 
the experimental bounds on the flavor-changing processes can provide 
stronger constraints on $m_{\rm rad}$ and $\Lmd_{\rd}$. 
According to the analysis of Ref.~\cite{Azatov:2008vm}, they should satisfy 
$\Lmd_{\rd}m_{\rm rad}>2.3a_{ds}\mbox{ TeV}^2$, 
where $a_{ds}$ is a dimensionless constant that parameterizes 
the flavor violation\footnote{
It is determined by the bulk mass parameters for the quark multiplets 
in the first two generations. }
and its typical values range between 0.03 and 0.12. 
Thus the most stringent bound on $m_{\rm rad}$ is that 
$m_{\rm rad}>2.7$~GeV for $e^{kL}=5\times 10^{13}$ and 
$m_{\rm rad}>10$~GeV for $e^{kL}=2.3\times 10^{14}$. 

\ignore{
For example, from the LEP data~\cite{Barate:2003sz}, 
the radion with a mass in the range:
$12\mbox{ GeV}<m_{\rm rad}<90\mbox{ GeV}$ is excluded 
if $(v/\Lmd_{\rd})^2>0.01$, 
which is translated into $e^{kL}>2.4\times 10^{14}$. 
This means that $\zeta=M_5/k$ must be larger than $1.3$ 
when $\sin\thH=0.05$ (see (\ref{def:zt})). 
Note that $m_{\rm rad}$, $m_{\rm H}$ and $\mKK$ depend on the warp factor 
only logarithmically, and does not change very much their values 
derived in the previous subsection. 
}

\section{Summary} \label{summary}
We have considered the modulus stabilization in the gauge-Higgs unification 
scenario, and estimate the radion and Higgs masses. 
Through the $\thH$-dependences of $M_5$ and $g_A$ in (\ref{thH-dependence}), 
various mass eigenvalues depend on $\thH$ nontrivially. 
We found that the masses of the radion, the Higgs boson 
and the KK modes all have different $\thH$-dependences. 
In order to see them explicitly, 
we considered two classes of models which correspond to different extensions 
of the model in Ref.~\cite{Hosotani:2008tx}. 
Qualitatively, we have the same results in both classes. 
\bea
 m_{\rm rad} \!\!\!&\sim\!\!\!& 
 \sqrt{\cO(20)}\times \frac{10^{-2}\mbox{ GeV}}{\sin^2\thH}, \nonumber\\
 m_{\rm H} \!\!\!&\sim\!\!\!& \sqrt{\cO(3)}\times 90\mbox{ GeV}, \nonumber\\
 \mKK \!\!\!&\sim\!\!\!& \frac{\cO(\mbox{500 GeV})}{\sin\thH}, 
\eea
and the radion-Higgs mixing is negligible. 
As mentioned in our previous work~\cite{Maru:2010ap}, 
the boundary kinetic terms for the gauge fields are necessary 
for the modulus stabilization. 
An $\cO(1)$ value of $\kp$ in (\ref{L_bd^kin}) is enough 
to stabilize the modulus. 
In contrast to the model in Ref.~\cite{Hosotani:2008tx}, 
the Higgs couplings to other particles do not deviate very much 
from the standard model values when $\thH\ll 1$. 

The radion couplings to the standard model particles are 
suppressed by factors of $\cO(v/\Lmd_{\rd})$ 
compared to the corresponding Higgs couplings. 
Within the allowed region of the parameter space, such suppression factors 
are less than 0.01. 
Therefore collider experiments do not impose any constraints on $m_{\rm rad}$. 
However the experimental bounds on the flavor-changing processes can provide 
stronger bounds on $m_{\rm rad}$ if the fermion mass hierarchy stems from 
the wavefunction localization. 
Such bounds narrow the allowed range of $\thH$ in some cases. 

Cosmological impacts of the radion physics is an intriguing subject. 
For this direction, we might need to extend the works 
by Refs.~\cite{Creminelli:2001th,Randall:2006py,Nardini:2007me} 
to deal with the one-loop effective potential at finite temperature 
in the Randall-Sundrum background. 
This is one of our future projects.

\vskip .5cm

\leftline{\bf Acknowledgments}
The author would like to thank N.~Maru for discussions 
in the early stage of this work.

\appendix

\section{Definitions of functions} \label{spectrum}
Here we define functions that are useful 
to express $\rho^I(\lmd;\thH)$ in (\ref{mass_det}) and 
the effective potential. 
First we define the following functions from the Bessel functions.  
\bea
 F_{\alp,\bt}^\kp(\lmd) \defa 
 J_\alp(\lmd)Y_\bt^\kp(\lmd z_L)-Y_\alp(\lmd)J_\bt^\kp(\lmd z_L), 
 \label{def:Fs}
\eea
where $z_L\equiv e^{kL}$, and 
\be
 J_\bt^\kp(u) \equiv J_\bt(u)-\kp u J_{\bt+1}(u), \;\;\;\;\;
 Y_\bt^\kp(u) \equiv Y_\bt(u)-\kp u Y_{\bt+1}(u). 
\ee
For calculations of the effective potential, we also define 
\be
 \hat{F}_{\alp,\bt}^\kp(w) \equiv I_\alp(w)K_\bt^\kp(w z_L)
 -e^{-i(\alp-\bt)\pi}K_\alp(w)I_\bt^\kp(w z_L), 
\ee
where
\be
 I_\bt^\kp(u) \equiv I_\bt(u)+\kp u I_{\bt+1}(u), \;\;\;\;\;
 K_\bt^\kp(u) \equiv K_\bt(u)-\kp u K_{\bt+1}(u).  \label{def:IK^kp}
\ee
Then the following relation holds. 
\be
 F_{\alp,\bt}^\kp(iw) = -\frac{2}{\pi}e^{i(\alp-\bt)\pi/2}
 \hat{F}_{\alp,\bt}^\kp(w). 
\ee

The asymptotic behavior of $\hat{F}^\kp_{\alp,\bt}(w)$ 
for $\Re w\gg 1$ is
\be
 \hat{F}^\kp_{\alp,\bt}(w) = -\frac{e^{w(z_L-1)}}{2w\sqrt{z_L}}
 e^{i(\bt-\alp)\pi}\brkt{1+\kp wz_L}\brc{1+\cO\brkt{w^{-1}}}. 
 \label{asp_hF}
\ee

\ignore{
For the gauge bosons, $\rho^I(\lmd)$ are given by 
\begin{description}
 \item{\bf Gluon sector}
 \be
  \rho^G(\lmd) = \lmd F_{0,0}^{\kp_c}(\lmd). 
 \ee
 \item{\bf W boson sector}
 \be
  \rho^W(\lmd) = F^{\kp_w}_{1,0}(\lmd)\brkt{F_{0,0}^{\kp_w}(\lmd)
 F_{1,1}^0(\lmd)-\frac{2\sin^2\thH}{\pi^2\lmd^2z_L}}.  
 \ee
 \item{\bf Neutral sector}
 \bea
 \rho^{\rm nt}(\lmd) \eql \cph^2\lmd F_{0,0}^{\kp_x}(\lmd)
 F_{1,0}^{\kp_w}(\lmd)\brkt{F_{0,0}^{\kp_w}(\lmd)F_{1,1}^0(\lmd)
 -\frac{2\sin^2\thH}{\pi^2\lmd^2z_L}} 
 \nonumber\\
 &&+\sph^2\lmd F_{1,0}^{\kp_x}(\lmd)F_{0,0}^{\kp_w}(\lmd)
 \brkt{F_{0,0}^{\kp_w}(\lmd)F_{1,1}^0(\lmd)
 -\frac{4\sin^2\thH}{\pi^2\lmd^2z_L}}. 
 \eea
 Especially, when $\kp_w=\kp_x$, the function~$\rho^{\rm nt}(\lmd)$ 
 can be factorized as $\rho^\gm(\lmd)\rho^Z(\lmd)$, 
 and the corresponding KK tower is decomposed into the following two sectors. 
 \begin{description}
  \item{\bf Photon sector}
  \be
   \rho^\gm(\lmd) = \lmd F_{0,0}^{\kp_w}(\lmd). 
  \ee
  \item{\bf Z boson sector}
  \be
   \rho^Z(\lmd) = F_{1,0}^{\kp_w}(\lmd)\brkt{
   F_{0,0}^{\kp_w}(\lmd)F_{1,1}^0(\lmd)
   -\frac{2(1+\sph^2)\sin^2\thH}{\pi^2\lmd^2z_L}}. 
  \ee
 \end{description}
 \item{\bf $\bdm{\hat{4}}$-component sector}
 \be
  \rho^{\hat{4}}(\lmd) = F_{1,0}^0(\lmd). 
 \ee
\end{description}
For quarks, there are four sectors according 
to the $\uem$ charge~$Q_{\rm EM}$. 
\begin{description}
 \item{\bf $\bdm{Q_{\rm EM}=\frac{5}{3}}$ sector}
 \be
  \rho^{5/3}(\lmd) = F_{c_1+\frac{1}{2},c_1-\frac{1}{2}}^0(\lmd). 
 \ee
 \item{\bf $\bdm{Q_{\rm EM}=\frac{2}{3}}$ sector}
 \bea
  \rho^{2/3}(\lmd) \eql F_{c_1+\frac{1}{2},c_1-\frac{1}{2}}^0\left\{
  s_\omg^2F_{c_2+\frac{1}{2},c_2-\frac{1}{2}}^0\brkt{
  F_{c_1+\frac{1}{2},c_1+\frac{1}{2}}^0F_{c_1-\frac{1}{2},c_1-\frac{1}{2}}^0
  -\frac{2\sin^2\thH}{\pi^2\lmd^2e^{kL}}} \right. \nonumber\\
  &&\left.\hspace{20mm}
  +c_\omg^2F_{c_1+\frac{1}{2},c_1+\frac{1}{2}}^0
  F_{c_1+\frac{1}{2},c_1-\frac{1}{2}}^0
  F_{c_2-\frac{1}{2},c_2-\frac{1}{2}}^0\right\}. 
 \eea
 \item{\bf $\bdm{Q_{\rm EM}=-\frac{1}{3}}$ sector}
 \bea
  \rho^{-1/3}(\lmd) \eql F_{c_2+\frac{1}{2},c_2-\frac{1}{2}}^0\left\{
  c_\omg^2F_{c_1+\frac{1}{2},c_1-\frac{1}{2}}^0\brkt{
  F_{c_2+\frac{1}{2},c_2+\frac{1}{2}}^0F_{c_2-\frac{1}{2},c_2-\frac{1}{2}}^0
  -\frac{2\sin^2\thH}{\pi^2\lmd^2e^{kL}}} \right. \nonumber\\
  &&\left.\hspace{20mm}
  +s_\omg^2F_{c_1-\frac{1}{2},c_1-\frac{1}{2}}^0
  F_{c_2+\frac{1}{2},c_2-\frac{1}{2}}^0
  F_{c_2+\frac{1}{2},c_2+\frac{1}{2}}^0\right\}. 
 \eea
 \item{\bf $\bdm{Q_{\rm EM}=-\frac{4}{3}}$ sector}
 \be
  \rho^{-4/3}(\lmd) = F_{c_2+\frac{1}{2},c_2-\frac{1}{2}}^0(\lmd). 
 \ee
\end{description}
The lepton sector has a similar structure to the quark sector. 
(See Ref.~\cite{Hosotani:2009qf}.) 
Besides, the KK tower of $\Psi_3$ consists of the $Q_{\rm EM}=\frac{2}{3}$ 
and $Q_{\rm EM}=-\frac{1}{3}$ sectors, and 
\bea
 \rho^{\Psi_3(2/3)}(\lmd) \eql \rho^{\Psi_3(-1/3)}(\lmd)
 = F_{c_3+\frac{1}{2},c_3+\frac{1}{2}}^0(\lmd)
 F_{c_3-\frac{1}{2},c_3-\frac{1}{2}}^0(\lmd)
 -\frac{4\cos^2\frac{\thH}{2}}{\pi^2\lmd^2e^{kL}}.  
 \label{rho_Psi3}
\eea
}

\section{Approximate form of effective potential} \label{ap:Veff}
Here we show the approximate forms of (\ref{expd:hatV}). 
As mentioned in Ref.~\cite{Garriga:2002vf}, 
the dominant contributions to the effective potential 
come from the gauge fields and the fermion fields with bulk masses 
that are close to $k/2$. 

As an example, let us consider a sector whose mass spectrum is determined by 
\be
 \rho^I(\lmd;\thH) = F_{\alp-1,\alp-1}^\kp(\lmd)F_{\alp,\alp}^0(\lmd)
 -\frac{2\sin^2\thH}{\pi^2\lmd^2e^{kL}} = 0, \label{eg:rho}
\ee
where $\alp$ is close to one. 
In fact, $\alp=1$ for the gauge sector, and $\alp=M_\Psi/k+\frac{1}{2}$ 
for the fermion sector. 
Then, the integrand of (\ref{def:hatV}) can be approximated for $\kL\gg 1$ as 
\bea
 \ln\frac{\rho^I(iwe^{-\kL};\thH)}{\cK^I(we^{-\kL})\cI^I(w)} 
 \eql \ln\frac{\rho^I(iwe^{-\kL};0)}{\cK^I(we^{-\kL})\cI^I(w)}
 +\ln\frac{\rho^I(iwe^{-\kL};\thH)}{\rho^I(iwe^{-\kL};0)} \nonumber\\
 \eql \ln\brc{1-\frac{I_{\alp-1}(we^{-\kL})K_{\alp-1}^\kp(w)}
 {K_{\alp-1}(we^{-\kp})I_{\alp-1}^\kp(w)}}
 +\ln\brc{1-\frac{I_\alp(we^{-\kL})K_\alp(w)}{K_\alp(we^{-\kL})I_\alp(w)}}
 \nonumber\\
 &&+\ln\brc{1+\frac{e^{\kL}\sin^2\thH}{2w^2\hat{F}_{\alp-1,\alp-1}^\kp(we^{-\kL})
 \hat{F}^0_{\alp,\alp}(we^{-\kL})}} \nonumber\\
 \sma -\frac{I_{\alp-1}(we^{-\kL})}{K_{\alp-1}(we^{-\kL})}
 \frac{K_{\alp-1}^\kp(w)}{I_{\alp-1}^\kp(w)}
 +\frac{e^{\kL}\sin^2\thH}{2w^2\hat{F}_{\alp-1,\alp-1}^\kp(we^{-\kL})
 \hat{F}_{\alp,\alp}^0(we^{-\kL})} \nonumber\\
 \sma \begin{cases} \frac{2(we^{-\kL}/2)^{2(\alp-1)}}
 {\Gm(\alp-1)\Gm(\alp)}\brc{-\frac{K_{\alp-1}^\kp(w)}{I_{\alp-1}^\kp(w)}
 +\frac{\sin^2\thH}{2wI_{\alp-1}^\kp(w)I_\alp(w)}}, & (\alp>1) \\
 -\frac{1}{\ln(we^{-\kL}/2)+\gm}\brc{-\frac{K_0^\kp(w)}{I_0^\kp(w)}
 +\frac{\sin^2\thH}{2wI_0^\kp(w)I_1(w)}}, & (\alp=1) \\
 \frac{2\sin(\pi\alp)}{\pi}\brc{-\frac{K_{\alp-1}^\kp(w)}{I_{\alp-1}^\kp(w)}
 +\frac{\sin^2\thH}{2wI_{\alp-1}^\kp(w)I_\alp(w)}}, & (\alp<1) \end{cases} 
 \label{eg:ln_rho}
\eea
where $kL$ in the definition of $\rho^I(\lmd;\thH)$ is replaced by $\kL$, 
and $\Gm(\alp)$ is the Gamma function. 
We have used that $\abs{\alp-1}\ll 1$, and assumed that $w=\cO(1)$ 
because the above function exponentially decays for $w\gg 1$.  
Thus the contribution of this sector to $\hat{V}(\kL,\thH)$ 
is negligible when $\alp>1$. 
When $\alp\leq 1$, it is estimated as 
\bea
 \hat{V}^I(\kL,\thH) \defa (-)^{2\eta_I}
 N_I\int_0^\infty\dr w\;w^3\ln\frac{\rho^I(iwe^{-\kL};\thH)}
 {\cK^I(we^{-\kL})\cI^I(w)} \nonumber\\
 \sma \begin{cases} \frac{(-)^{2\eta_I}}{\kL}
 \sum_{n=0}^2v_n^I\cos^n\thH, & (\alp=1) \\
 (-)^{2\eta_I}\frac{2\sin(\pi\alp)}{\pi}
 \sum_{n=0}^2v_n^I\cos^n\thH, & (\alp<1) \end{cases}
\eea
where
\bea
 v_0^I \eql N_I\int_0^\infty\dr w\;w^3\brc{-\frac{K_{\alp-1}^\kp(w)}
 {I_{\alp-1}^\kp(w)}+\frac{1}{2wI_{\alp-1}^\kp(w)I_\alp(w)}}, \nonumber\\
 v_1^I \eql 0, \;\;\;\;\;
 v_2^I = -N_I\int_0^\infty\dr w\;\frac{w^2}{2I_{\alp-1}^\kp(w)I_\alp(w)}. 
\eea
Here we have used that $\vph\gg\abs{\ln(w/2)+\gm}$ in the integration region 
that gives dominant contributions. 
For $\kp=1.0$ and $\alp=1$, for example, $v_0^I\simeq 8.6$ 
and $v_2^I\simeq -5.4$. 
In general, we can see that $v_0^I=\cO(N_I)$ 
and $v_2^I=\cO(-N_I)$.\footnote{
For the fermion sector, the forms of $\rho^I(\lmd;\thH)$ are more complicated 
than (\ref{eg:rho}), but the above rough estimate does not change much. 
} 
For the gluon and photon sectors, in which 
$\rho^{I=G,\gm}(\lmd;\thH)=\lmd F_{0,0}^\kp(\lmd)$, 
the terms proportional to $\sin^2\thH$ in (\ref{eg:ln_rho}) are absent. 
Thus we find that $v_0^I=-\cO(N_I)$ and $v_1^I=v_2^I=0$, 
where $N_G=24$ and $N_\gm=3$.  

For the $\Psi_3$-sector, in which $\rho^I(\lmd;\thH)$ is given 
by (\ref{rho_Psi3}), 
we can estimate $\hat{V}^{I=\Psi_3}(\kL,\thH)$ in a similar way and find that 
$v_0^{\Psi_3}=\cO(N_{\Psi_3})=\cO(24)$, $v_1^{\Psi_3}=\cO(-24)$ 
and $v_{n=2}^{\Psi_3}=0$. 

As a result, we obtain 
\bea
 \hat{V}(\kL,\thH) \eql \sum_I\hat{V}^I(\kL,\thH) \nonumber\\
 \sma -\frac{2\sin(\pi\alp_1)}{\pi}\brkt{v_0^{\Psi_1}+v_2^{\Psi_1}\cos^2\thH}
 -\frac{2\sin(\pi\alp_3)}{\pi}\brkt{v_0^{\Psi_3}+v_1^{\Psi_3}\cos\thH} 
 \nonumber\\
 &&+\sum_{I=W,Z}\frac{1}{\kL}\brkt{v_0^I+v_2^I\cos^2\thH}
 +\sum_{I=G,\gm}\frac{v_0^I}{\kL} \nonumber\\
 \eql \sum_{n=0}^2\brkt{u_n+\frac{v_n}{\kL}}\cos^n\thH,  \label{ap:hatV}
\eea
where $u_0=\cO(-5)$, $u_1=u_2=\cO(3)$, $v_0=v_2=\cO(10)$ 
and $v_1=0$. 

\ignore{
The contributions from the gauge fields contain 
the ratio~$\frac{I_0(we^{-\kL})}{K_0(we^{-\kL})}$, 
in the integrand of (\ref{def:hatV}), which is approximated 
for $we^{-\kL}\ll 1$ as 
$\frac{I_0(we^{-\kL})}{K_\alp(we^{-\kL})} 
\simeq -\brc{\ln(we^{-\kL}/2)+\gm}^{-1}$, 
where $\gm$ is the Euler's constant. 
As an example, let us consider the W boson sector. 
Since $\rho^W(\lmd;\thH)$ is given by 
\be
 \rho^W(\lmd;\thH) = F_{1,0}^\kp(\lmd)
 \brkt{F_{0,0}^\kp(\lmd)F_{1,1}^0(\lmd)
 -\frac{2\sin^2\thH}{\pi^2\lmd^2e^{kL}}},  \label{rhoW}
\ee
the integrand in (\ref{def:hatV}) for this sector 
can be approximated for $\kL\gg 1$ as 
\bea
 &&\ln\frac{\rho^W(iwe^{-\kL};\thH)}{\cK^W(we^{-\kL})\cI^W(w)} 
 = \ln\frac{\rho^W(iwe^{-\kL};0)}{\cK^W(we^{-\kL})\cI^W(w)}
 +\ln\frac{\rho^W(iwe^{-\kL};\thH)}{\rho^W(iwe^{-\kL};0)} \nonumber\\
 \eql \ln\brc{1-\frac{I_1(we^{-\kL})K_0^\kp(w)}{K_1(we^{-\kL})I_0^\kp(w)}}
 +\ln\brc{1-\frac{I_0(we^{-\kL})K_0^\kp(w)}{K_0(we^{-\kL})I_0^\kp(w)}} 
 \nonumber\\
 &&+\ln\brc{1-\frac{I_1(we^{-\kL})K_1^\kp(w)}{K_1(we^{-\kL})I_1^\kp(w)}} 
 +\ln\brc{1+\frac{e^{\kL}\sin^2\thH}
 {2w^2\hat{F}_{0,0}^\kp(we^{-\kL})\hat{F}_{1,1}^0(we^{-\kL})}} \nonumber\\
 \sma -\frac{I_0(we^{-\kL})}{K_0(we^{-\kL})}\frac{K_0^\kp(w)}{I_0^\kp(w)}
 +\frac{e^{\kL}\sin^2\thH}
 {2w^2\hat{F}_{0,0}^\kp(we^{-\kL})\hat{F}_{1,1}^0(we^{-\kL})} \nonumber\\
 \sma \frac{1}{\ln(we^{-\kL}/2)+\gm}\brc{\frac{K_0^\kp(w)}{I_0^\kp(w)}
 -\frac{\sin^2\thH}{2wI_0^\kp(w)I_1(w)}},  
\eea
where $kL$ in the definition of $\rho^W(\lmd;\thH)$ is replaced by $\kL$. 
We assumed that $w=\cO(1)$ 
because the above function exponentially decays for $w\gg 1$. 
Therefore, the contribution of this sector to $\hat{V}(\kL,\thH)$ is 
estimated as 
\bea
 \hat{V}^W(\kL,\thH) \!\!\!&\sim\!\!\!& 6\int_0^\infty\dr w\;w^3
 \sbk{\frac{1}{\ln(we^{-\kL}/2)+\gm}\brc{\frac{K_0^\kp(w)}{I_0^\kp(w)}
 -\frac{\sin^2\thH}{2wI_0^\kp(w)I_1(w)}}} \nonumber\\
 \sma \frac{v_0^W+v_2^W\cos^2\thH}{\kL}, 
\eea
where 
\bea
 v_0^W \defa -6\int_0^\infty\dr w\;w^3\brc{\frac{K_0^\kp(w)}{I_0^\kp(w)}
 -\frac{1}{2wI_0^\kp(w)I_1(w)}}, \nonumber\\
 v_2^W \defa -3\int_0^\infty\dr w\;\frac{w^2}{I_0^\kp(w)I_1(w)}. 
\eea
Here we have used that $\kL\gg \abs{\ln(w/2)+\gm}$ 
in the integration region that gives dominant contributions. 
When $\kp=1.0$, these are estimated as $v_0^W\simeq 8.6$ and 
$v_2^W\simeq -5.4$. 
The contributions of the other gauge sectors are similarly estimated, and 
$v_n^I=\cO(N_I)$ are obtained. 
}

\section{Expressions for numerical calculations} \label{expr_for_cal}
Although the approximate expressions of the mass eigenvalues in 
(\ref{ap:m_scalar2}) or (\ref{ap:m_scalar3}) is useful 
for the order estimation of the mass eigenvalues, 
we need more accurate expression for the numerical calculation. 

Here we consider the case of (\ref{ap:Veff2}) as an example. 
Then, since the functions~$\rho^I(\lmd;\thH)$ have the form of 
\be
 \rho^I(\lmd;\thH) = \rho_0^I(\lmd)+\rho_1^I(\lmd)\cos\thH
 +\rho_2^I(\lmd)\cos^2\thH, 
\ee
we can expand the integrand of (\ref{def:hatV}) 
around $\cos\thH=c_0$ ($c_0$ is a constant) as 
\bea
 \ln\rho^I \eql \ln\brc{\rho_0^I+\rho_1^Ic_0+\rho_2^Ic_0^2}
 -\sgm_1^Ic_0+\sgm_2^Ic_0^2 \nonumber\\
 &&+\brkt{\sgm_1^I-2\sgm_2^Ic_0}\cos\thH
 +\sgm_2^I\cos^2\thH+\cO\brkt{(\cos\thH-c_0)^3},  \label{expd:rho}
\eea
where 
\be
 \sgm_1^I \equiv \frac{\rho_1^I+2\rho_2^Ic_0}
 {\rho_0^I+\rho_1^Ic_0+\rho_2^Ic_0^2}, \;\;\;\;\;
 \sgm_2^I \equiv \frac{2\rho_0^I\rho_2^I-\brkt{\rho_1^I}^2
 -2\rho_1^I\rho_2^Ic_0-2\brkt{\rho_2^I}^2c_0^2}
 {2\brkt{\rho_0^I+\rho_1^Ic_0+\rho_2^Ic_0^2}^2}. 
\ee
Thus $\hat{V}(\kL,\thH)$ in (\ref{def:hatV}) is expanded as 
\be
 \hat{V}(\kL,\thH) = \hat{V}_0(\kL;c_0)+\hat{V}_1(\kL;c_0)\cos\thH
 +\hat{V}_2(\kL;c_0)\cos^2\thH+\cO\brkt{(\cos\thH-c_0)^3}, 
 \label{exact_expd}
\ee
where 
\bea
 \hat{V}_0(\kL;c_0) \eql \sum_I(-)^{2\eta_I}N_I\int_0^\infty\dr w \;w^3
 \left\{\ln\frac{\brkt{\rho_0^I+\rho_1^Ic_0+\rho_2^Ic_0^2}(iwe^{-\kL})}
 {\cK^I(we^{-\kL})\cI^I(w)} \right. \nonumber\\
 &&\left.\hspace{45mm}
 -\brkt{\sgm_1^Ic_0-\sgm_2^Ic_0^2}(iwe^{-\kL})\right\}, \nonumber\\
 \hat{V}_1(\kL;c_0) \eql \sum_I(-)^{2\eta_I}N_I\int_0^\infty\dr w\;w^3
 \brkt{\sgm_1^I-2\sgm_2^Ic_0}(iwe^{-\kL}), \nonumber\\
 \hat{V}_2(\kL;c_0) \eql \sum_I(-)^{2\eta_I}N_I
 \int_0^\infty\dr w\;w^3\sgm_2^I(iwe^{-\kL}). 
\eea
If we choose a constant~$c_0$ as a solution to the equation,  
\be
 c_0 = -\frac{\hat{V}_1(kL;c_0)}{\hat{V}_2(kL;c_0)}, 
\ee
the correction term~$\cO((\cos\thH-c_0)^3)$ in (\ref{exact_expd}) 
can be neglected in the calculations of $m_{\rm rad}$ and $m_{\rm H}$.



\begin{thebibliography}{99}
 \bibitem{Fairlie:1979at}
  D.~B.~Fairlie,
  Phys.\ Lett.\  B {\bf 82} (1979) 97; 
  N.~S.~Manton,
  Nucl.\ Phys.\  B {\bf 158} (1979) 141.

 \bibitem{Hosotani:1983xw}
  Y.~Hosotani,
  Phys.\ Lett.\  B {\bf 126} (1983) 309; 
  Y.~Hosotani,
  Annals Phys.\  {\bf 190} (1989) 233.

 \bibitem{Hatanaka:1998yp}
  H.~Hatanaka, T.~Inami and C.~S.~Lim,
  Mod.\ Phys.\ Lett.\  A {\bf 13} (1998) 2601; 

 \bibitem{Pomarol:1998sd}
  A.~Pomarol and M.~Quiros,
  Phys.\ Lett.\  B {\bf 438} (1998) 255.

 \bibitem{Antoniadis:2001cv}
  I.~Antoniadis, K.~Benakli and M.~Quiros,
  New J.\ Phys.\  {\bf 3} (2001) 20; 
  G.~von Gersdorff, N.~Irges and M.~Quiros,
  Nucl.\ Phys.\  B {\bf 635} (2002) 127; 
  C.~S.~Lim, N.~Maru and K.~Hasegawa,
  J.\ Phys.\ Soc.\ Jap.\  {\bf 77} (2008) 074101; 
  N.~Maru and T.~Yamashita,
  Nucl.\ Phys.\  B {\bf 754} (2006) 127; 
  Y.~Hosotani, N.~Maru, K.~Takenaga and T.~Yamashita,
  Prog.\ Theor.\ Phys.\  {\bf 118} (2007) 1053. 

 \bibitem{Csaki:2002ur}
  C.~Csaki, C.~Grojean and H.~Murayama,
  Phys.\ Rev.\  D {\bf 67} (2003) 085012; 
  C.~A.~Scrucca, M.~Serone and L.~Silvestrini,
  Nucl.\ Phys.\  B {\bf 669} (2003) 128. 

 \bibitem{Hall:2001zb}
  L.~J.~Hall, Y.~Nomura and D.~Tucker-Smith,
  Nucl.\ Phys.\  B {\bf 639} (2002) 307;  
  L.~J.~Hall, H.~Murayama and Y.~Nomura,
  Nucl.\ Phys.\  B {\bf 645} (2002) 85. 

 \bibitem{Randall:1999ee}
  L.~Randall and R.~Sundrum,
  Phys.\ Rev.\ Lett.\  {\bf 83} (1999) 3370. 

 \bibitem{Hosotani:2005nz}
  Y.~Hosotani and M.~Mabe,
  Phys.\ Lett.\  B {\bf 615} (2005) 257. 

 \bibitem{Agashe:2004rs}
  K.~Agashe, R.~Contino and A.~Pomarol,
  Nucl.\ Phys.\  B {\bf 719} (2005) 165. 

\bibitem{Contino:2006qr}
  R.~Contino, L.~Da Rold and A.~Pomarol,
  Phys.\ Rev.\  D {\bf 75}, 055014 (2007).

 \bibitem{Hosotani:2006qp}
  Y.~Hosotani, S.~Noda, Y.~Sakamura and S.~Shimasaki,
  Phys.\ Rev.\  D {\bf 73} (2006) 096006. 

 \bibitem{Hosotani:2007qw}
  Y.~Hosotani and Y.~Sakamura,
  Phys.\ Lett.\  B {\bf 645} (2007) 442; 
  Prog.\ Theor.\ Phys.\  {\bf 118} (2007) 935; 
  Y.~Sakamura,
  Phys.\ Rev.\  D {\bf 76} (2007) 065002. 

 \bibitem{Hosotani:2008tx}
  Y.~Hosotani, K.~Oda, T.~Ohnuma and Y.~Sakamura,
  Phys.\ Rev.\  D {\bf 78}, 096002 (2008)
  [Erratum-ibid.\  D {\bf 79}, 079902 (2009)]. 

 \bibitem{Haba:2009ei}
  N.~Haba, Y.~Sakamura and T.~Yamashita,
  JHEP {\bf 0907} (2009) 020; 
  arXiv:0908.1042 [hep-ph].

 \bibitem{Hosotani:2009jk}
  Y.~Hosotani, P.~Ko and M.~Tanaka,
  Phys.\ Lett.\  B {\bf 680} (2009) 179. 

 \bibitem{Goldberger:1999uk}
  W.~D.~Goldberger and M.~B.~Wise,
  Phys.\ Rev.\ Lett.\  {\bf 83} (1999) 4922. 

 \bibitem{Fabinger:2000jd}
  M.~Fabinger and P.~Horava,
  Nucl.\ Phys.\  B {\bf 580} (2000) 243. 

 \bibitem{Garriga:2000jb}
  J.~Garriga, O.~Pujolas and T.~Tanaka,
  Nucl.\ Phys.\  B {\bf 605} (2001) 192;  
  D.~J.~Toms,
  Phys.\ Lett.\  B {\bf 484} (2000) 149;  
  W.~D.~Goldberger and I.~Z.~Rothstein,
  Phys.\ Lett.\  B {\bf 491} (2000) 339; 
  I.~H.~Brevik, K.~A.~Milton, S.~Nojiri and S.~D.~Odintsov,
  Nucl.\ Phys.\  B {\bf 599} (2001) 305. 

 \bibitem{Hofmann:2000cj}
  R.~Hofmann, P.~Kanti and M.~Pospelov,
  Phys.\ Rev.\  D {\bf 63} (2001) 124020. 

 \bibitem{Ponton:2001hq}
  E.~Ponton and E.~Poppitz,
  JHEP {\bf 0106} (2001) 019. 

 \bibitem{Garriga:2002vf}
  J.~Garriga and A.~Pomarol,
  Phys.\ Lett.\  B {\bf 560} (2003) 91. 

\bibitem{Maru:2010ap}
  N.~Maru and Y.~Sakamura,
  JHEP {\bf 1004}, 100 (2010)

 \bibitem{Cacciapaglia:2005da}
  G.~Cacciapaglia, C.~Csaki and S.~C.~Park,
  JHEP {\bf 0603} (2006) 099. 

 \bibitem{Medina:2007hz}
  A.~D.~Medina, N.~R.~Shah and C.~E.~M.~Wagner,
  Phys.\ Rev.\  D {\bf 76} (2007) 095010. 
  
 \bibitem{Davoudiasl:1999tf}
  H.~Davoudiasl, J.~L.~Hewett and T.~G.~Rizzo,
  Phys.\ Lett.\  B {\bf 473}, 43 (2000). 

 \bibitem{Giudice:2000av}
  G.~F.~Giudice, R.~Rattazzi and J.~D.~Wells,
  Nucl.\ Phys.\  B {\bf 595}, 250 (2001). 

\bibitem{Csaki:2000zn}
  C.~Csaki, M.~L.~Graesser and G.~D.~Kribs,
  Phys.\ Rev.\  D {\bf 63}, 065002 (2001). 
  
\bibitem{Davoudiasl:2002ua}
  H.~Davoudiasl, J.~L.~Hewett and T.~G.~Rizzo,
  Phys.\ Rev.\  D {\bf 68} (2003) 045002
  [arXiv:hep-ph/0212279].
  
\bibitem{:2010ar}
    [CDF and D0 Collaboration],
  arXiv:1007.4587 [hep-ex].

\bibitem{Csaki:1999mp}
  C.~Csaki, M.~Graesser, L.~Randall and J.~Terning,
  Phys.\ Rev.\  D {\bf 62}, 045015 (2000). 

\bibitem{Goldberger:1999un}
  W.~D.~Goldberger and M.~B.~Wise,
  Phys.\ Lett.\  B {\bf 475}, 275 (2000). 
  
\bibitem{Rizzo:2002pq}
  T.~G.~Rizzo,
  JHEP {\bf 0206}, 056 (2002). 


\bibitem{Csaki:2007ns}
  C.~Csaki, J.~Hubisz and S.~J.~Lee,
  Phys.\ Rev.\  D {\bf 76}, 125015 (2007). 
  
\bibitem{Azatov:2008vm}
  A.~Azatov, M.~Toharia and L.~Zhu,
  Phys.\ Rev.\  D {\bf 80}, 031701 (2009). 


\bibitem{Barate:2003sz}
  R.~Barate {\it et al.}  [LEP Working Group for Higgs boson searches and
                  ALEPH Collaboration and  and],
  Phys.\ Lett.\  B {\bf 565}, 61 (2003).
  
\bibitem{Acton:1991pd}
  P.~D.~Acton {\it et al.}  [OPAL Collaboration],
  Phys.\ Lett.\  B {\bf 268}, 122 (1991).

\bibitem{Mahanta:2000mx}
  U.~Mahanta and S.~Mohanty,
  Phys.\ Rev.\  D {\bf 62}, 083003 (2000).

\bibitem{Falkowski:2007hz}
  A.~Falkowski,
  Phys.\ Rev.\  D {\bf 77}, 055018 (2008). 

\bibitem{Maru:2007xn}
  N.~Maru and N.~Okada,
  Phys.\ Rev.\  D {\bf 77}, 055010 (2008). 

\bibitem{Goldberger:2007zk}
  W.~D.~Goldberger, B.~Grinstein and W.~Skiba,
  Phys.\ Rev.\ Lett.\  {\bf 100}, 111802 (2008). 



\bibitem{Creminelli:2001th}
  P.~Creminelli, A.~Nicolis and R.~Rattazzi,
  JHEP {\bf 0203}, 051 (2002). 
  
 \bibitem{Randall:2006py}
  L.~Randall and G.~Servant,
  JHEP {\bf 0705}, 054 (2007). 


\bibitem{Nardini:2007me}
  G.~Nardini, M.~Quiros and A.~Wulzer,
  JHEP {\bf 0709} (2007) 077. 

\end{thebibliography}
\end{document}